\documentclass[12pt]{iopart}
\usepackage[english]{babel}
\usepackage{graphicx}
\usepackage{bm}
\usepackage{color}

\begin{document}

\title{Swift heavy ion irradiation of SrTiO$_3$ under grazing incidence}

\author{Ender Akc\"oltekin$^1$, Sevilay Akc\"oltekin$^1$, Orkhan Osmani$^1$, Henning Lebius$^2$,
Marika Schleberger$^1$\footnote{electronic address:marika.schleberger@uni-due.de}}
\address{$^1$ Experimentelle Physik, Universit\"at Duisburg-Essen, 47048 Duisburg, Germany}
\address{$^2$ CIMAP, blvd. Henri Becquerel, 14070 Caen Cedex 5, France}



\begin{abstract}The irradiation of SrTiO$_3$ single crystals with
swift heavy ions leads to modifications of the surface. The
details of the morphology of these modifications depends strongly
on the angle of incidence and can be characterized by atomic
force microscopy. At glancing angles, discontinuous chains of
nanosized hillocks appear on the surface. The latent track radius
can be determined from the variation of the length of the chains
with the angle of incidence. This radius is material specific and
allows the calculation of the electron-phonon-coupling constant
for SrTiO$_3$. We show that a theoretical description of the
nanodot creation is possible within a two-temperature model if the
spatial electron density is taken into account. The appearance of
discontinuous features can be explained easily within this model,
but it turns out that the electronic excitation dissipates on a
femtosecond time scale and thus too rapidly to feed sufficient
energy into the phonon system in order to induce a thermal
melting process. We demonstrate that this can be solved if the
temperature dependent diffusion coefficient is introduced into
the model.
\end{abstract}


\maketitle


\section{Introduction}
The irradiation of solid matter with heavy ions of 100 MeV energy
has long been known to create structural modifications ranging
from defects and amorphization in the bulk up to the creation of
hillocks on the surface
\cite{Awazu2006,Khalfaoui2006,El-Said2004,Mueller1998}.

In this energy range the projectile is slowed down almost
exclusively by electronic interactions. The standard model to
describe the energy transport is based on a two-temperature model
and includes solving the heat diffusion equation in a cylindrical
symmetry. This approach requires many approximations as well as
the fitting of the strength of the electron-phonon coupling. The
model has been quite successful explaining e.g.~track radii in
various materials created by irradiation perpendicular with
respect to the surface \cite{Toulemonde1992}.

If the sample is irradiated under grazing incidence it has
recently been shown that chains of nanodots are created by single
ions \cite{Akcoeltekin2007,Carvalho2007}. In an earlier
experiment similar elongated tracks on a LiF surface covered with
gold particles have been observed and were explained in terms of
an elastic shock wave model \cite{Vorobyova1998}. These phenomena
cannot be longer described by the conventional thermal spike
model for several reasons. ? the electron density cannot be
approximated by a free electron gas because the electronic
structure plays a major role in the creation of the chains.

To explain these striking experimental findings and to overcome
the limitations of the conventional approach a model was proposed
where the creation of hillocks is directly coupled to the spatial
electron density of the material \cite{Akcoeltekin2007}. Every
time the projectile travels through a region with a high density
the energy loss is sufficiently high to feed energy into the
electronic system, which finally results in a nanodot on the
surface. Since the electron density corresponds to the periodicity
of the crystal the nanodots appear on the surface with a certain
periodicity depending on the exact impact parameters.

The aim of this paper is to study the track creation in SrTiO$_3$
in more detail. We extend the model proposed by Akc\"oltekin et
al.~\cite{Akcoeltekin2007} in order to determine the
electron-phonon coupling constant and to study the diffusion of
energy within the electron and the phonon system.

\section{Experimental procedures} The single crystal samples of
SrTiO$_3$(100) and SrTiO$_3$(111)(Crystec, Berlin) have been
irradiated without prior surface treatment at the ion beam
facility IRRSUD of the GANIL, France. The irradiation was done
using $^{207}$Pb$^{28+}$ and $^{131}$Xe$^{23+}$ ions with kinetic
energies of 130 and 93~MeV, respectively. These energies
correspond to stopping powers of 22~keV/nm and 19~keV/nm,
respectively, as calculated with SRIM \cite{srim}. The angle of
incidence with respect to the surface was varied between
$\Theta=90^\circ$ and $\Theta=0.3^\circ$. Fluences were typically
chosen to yield between 5 and 20 impacts per $\mu$m$^2$. After
irradiation the samples were analyzed by atomic force microscopy
(AFM) either in the intermittent contact mode under ambient
conditions or in contact mode in UHV. All AFM images were
processed with the Nanotec Electronica SL WSxM software, version
4.0 Develop 8.3. \cite{wsxm}. From the raw data (400 $\times$ 400
data points) only a plane was subtracted. The colour code was
changed using the palette {\it flow.lut}. No change of contrast
(1) or brightness (0) was used.

\section{Experimental results}
\begin{figure}
\includegraphics[width=4cm]{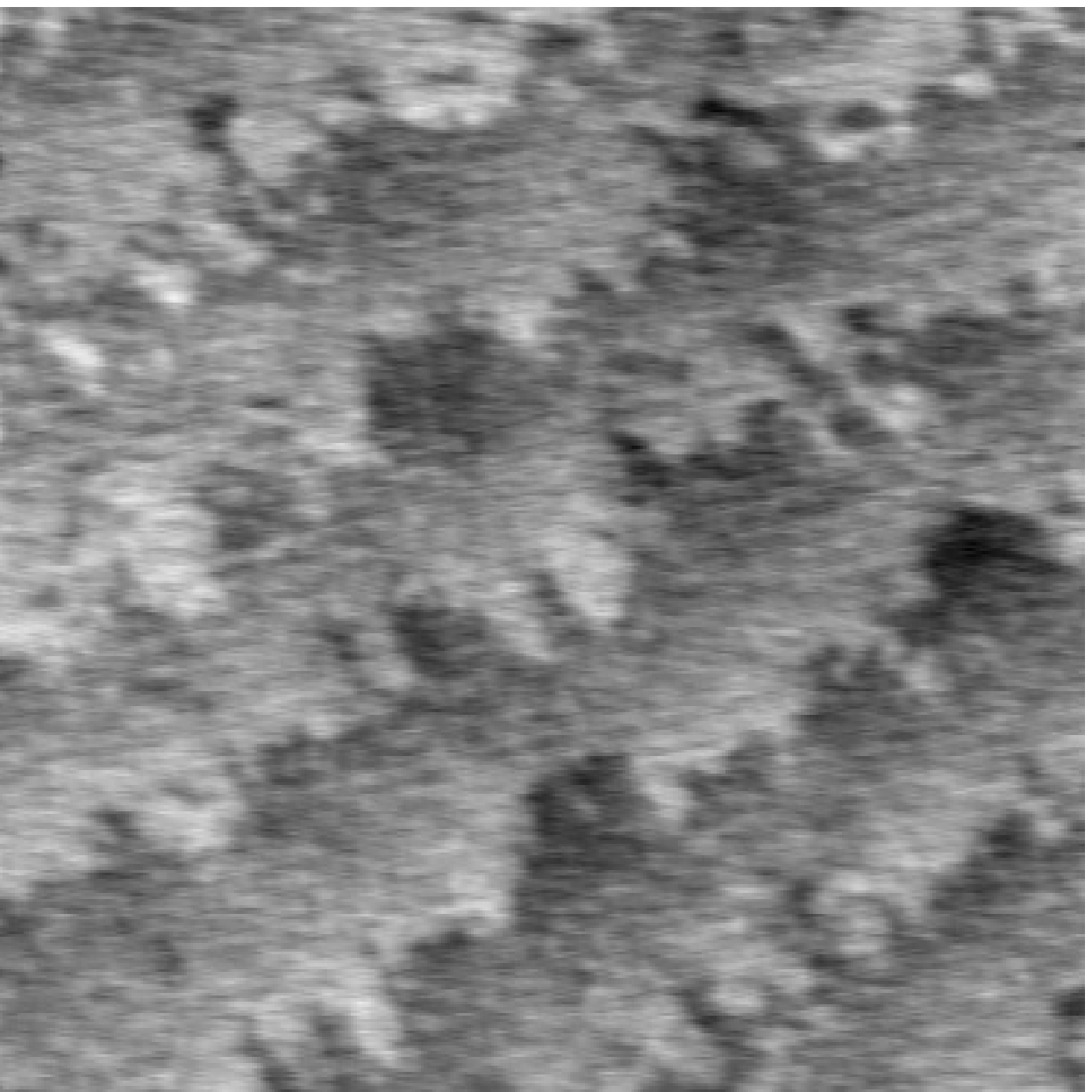}
\hspace{1cm}
\includegraphics[width=4cm]{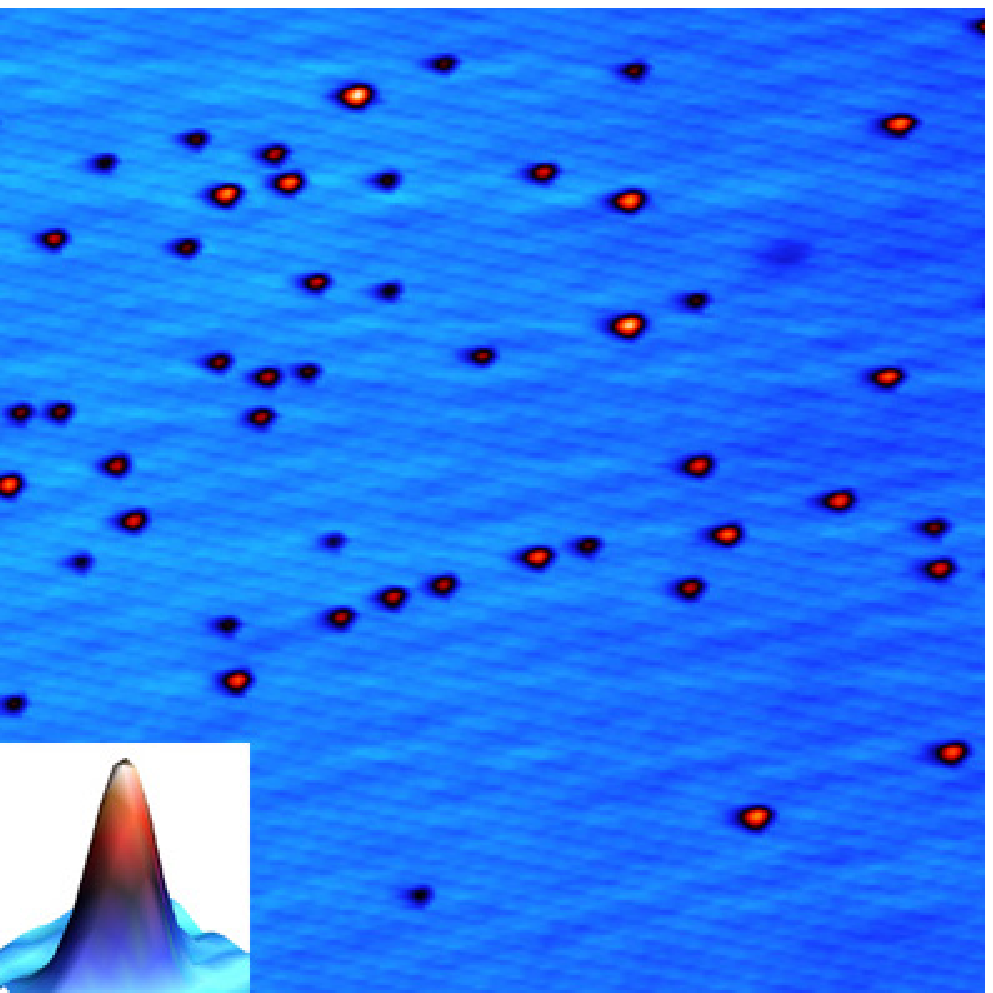}
\caption{Left panel: AFM image of a clean SrTiO$_3$(100) surface
taken {\it in situ}, roughness is $rms <$1~nm. Frame size is 800
$\times$ 800 nm$^2$. Right panel: Surface after irradiation with
130~MeV Pb ions, $\Theta=90^\circ$, fluence 3 $\times
10^9$~ions/cm$^2$. Inset shows a 3-dimensional image of a single
hillock which has a height of 4.5 nm. Image size is $2 \times
2~\mu$m$^2$.} \label{sauber}
\end{figure}

Since the samples undergo no surface treatment prior to
irradiation we make sure that the virgin surface is sufficiently
characterized to be able to identify modifications due to
irradiation. In the left part of fig.\ref{sauber} a typical AFM
image of an untreated SrTiO$_3$ surface is shown. The irregular
step edges separating the terraces can be clearly seen. The
average step height is about $(3\pm0.5)$~\AA, this is in
reasonable agreement with the literature value of 3.905~\AA
\cite{handbook}. The surface roughness is $rms \leq 1$~nm.


After irradiation at $\Theta=90^\circ$ nanosized hillocks have
appeared on the otherwise unchanged surface. The sample shown in
the right part of fig.\ref{sauber} was exposed to a fluence of $1
\times 10^9$~ions/cm$^2$. On average we find 11 hillocks on this
surface, i.e.~every ion impact produces one of these nanodots.

\begin{figure}
\includegraphics[width=15cm]{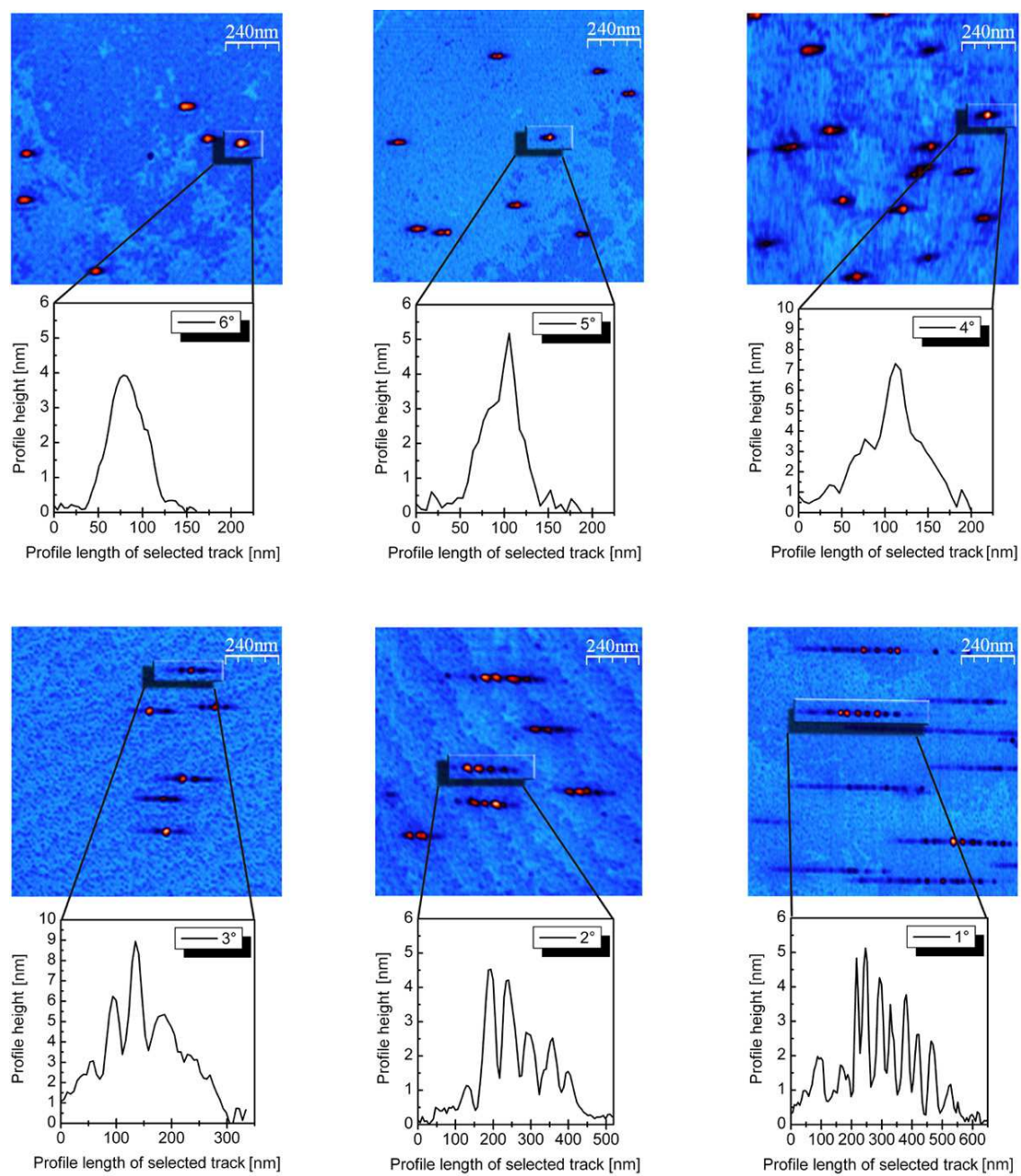}
\caption{AFM images of SrTiO$_3$ surfaces irradiated with 93~MeV
Xe ions. Upper row, from left to right: angle of incidence
$6^\circ,5^\circ,4^\circ$. Lower row, from left to right. angle
of incidence $3^\circ, 2^\circ,1^\circ$. Image size is $1.2
\times 1.2~\mu$m$^2$, scale runs from 0~nm (blue) to 5~nm
(orange). Line scans are taken from highlighted regions.}
\label{winkelserie}
\end{figure}

If we change the angle of incidence from $\Theta=90^\circ$ down to
angles of 10$^\circ$, the shape of the hillocks remains unchanged.
However, if the angle is decreased even further the morphology of
the irradiation induced damage changes drastically. This can be
seen from fig.\ref{winkelserie}. This series of AFM images
demonstrates that the hillocks at first elongate along the
direction of the ion trajectory (indicated by arrows), then
elongated structures with two or three separate maxima appear,
and finally chains of totally disconnected hillocks are created.
Again, the number of chains corresponds to the fluence, so that
each ion creates one chain. These chains can have a total length
of a few microns, contain dozens of nanodots and these nanodots
seem to exhibit a certain periodicity. The dimensions of the
nanodots within the chains are comparable to the dimensions
determined from the $\Theta=90^\circ$ experiment (see insets).


By comparing samples irradiated with xenon ions with those
irradiated with lead ions (not shown here), no significant
differences in the morphology could be detected.

\section{Discussion}
\begin{figure}
\includegraphics[width=7cm]{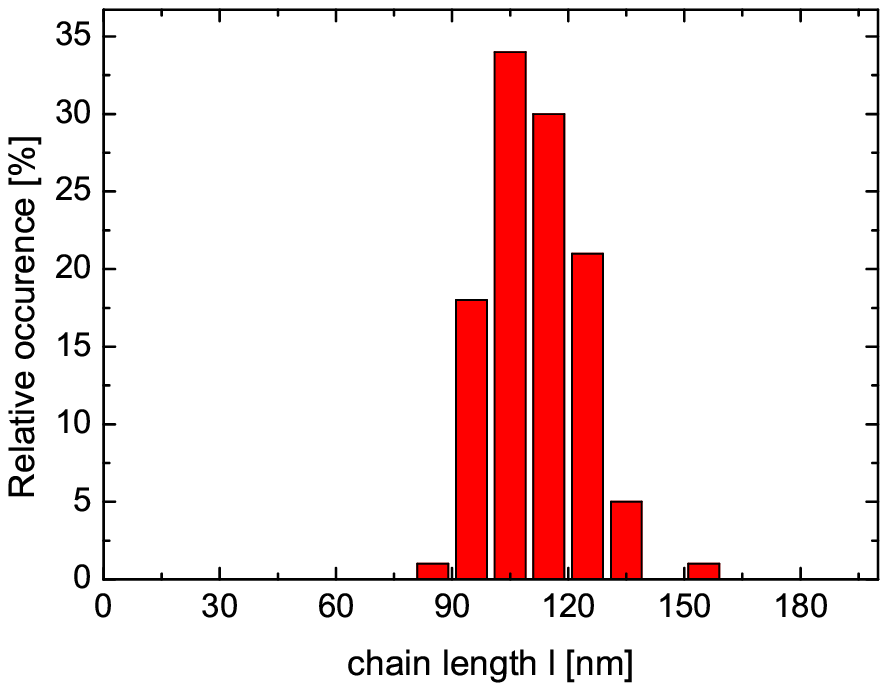}
\hspace{1cm}
\includegraphics[width=7cm]{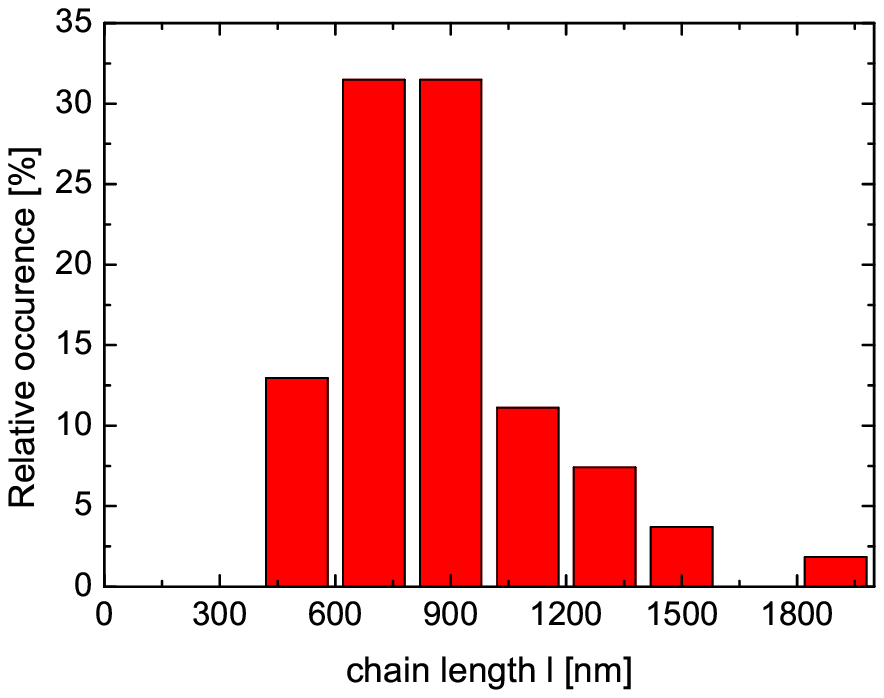}
\caption{Length distribution for two different angles of
incidence ($\Theta=5^\circ$ and $\Theta=0.5^\circ$) determined
from 105 and 60 individual tracks, respectively.}
\label{verteilung}
\end{figure}

It is clearly apparent from fig.\ref{winkelserie} that the length
of the track varies with the angle of incidence. In order to
quantify this we determined the length distributions as a
function of the angle of incidence by analyzing around 60 tracks
for each angle. As an example we present the length distribution
for $\Theta=0.5^\circ$ and $\Theta=5^\circ$, respectively (see
fig.\ref{verteilung}). Both are Gauss-shaped, for the extremely
glancing angle it is centered around 750~nm whereas the mean
length at 5$^\circ$ is 100~nm. The distribution for the glancing
angels is much broader, most likely due to surface properties and
the exact impact point (see also below).

\begin{figure}
\includegraphics[width=10cm]{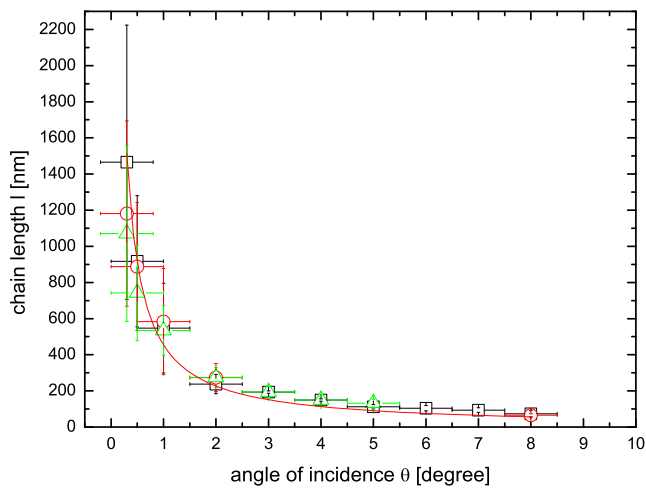}
\caption{Length of chains as a function of angle of incidence.
Data from various experiments. Circles: 96 MeV Xe$^{23+}$ $\rightarrow$
SrTiO$_3$(111). Squares: 96 MeV Xe$^{23+}$ $\rightarrow$ SrTiO$_3$(100).
Triangles: 130 MeV Pb$^{28+}$ $\rightarrow$ SrTiO$_3$(100). The line
represents a fit according to eq.~\ref{eq1} with $d=8$~nm.}
\label{winkel}
\end{figure}

The geometric relation describing the angle dependence has
already been established in \cite{Akcoeltekin2007} and is applied
here as well (see fig.\ref{winkel}):
\begin{equation}
l=d/\tan \Theta \label{eq1}
\end{equation}
with the chain length $l$, and $d$ being the maximum depth from
which the excitation may reach the surface. This simple model
assumes that a part of the latent track in the volume can be
detected at the surface as a chain of hillocks. The ion excites
the electronic system along the track and the energy is later
transferred to the lattice via electron-phonon coupling. In a
region where the local energy density is high enough, defects can
be created, melting and amorphization may occur. Therefore, the
ions trajectory is surrounded by a modified region with a radius
given by $d$.

How far this local excitation can spread (i.e. the value of $d$)
depends entirely on the material properties, such as the
electron-phonon-coupling constant, the heat capacity, the heat
conductivity, and the melting temperature. In a metal e.g., it is
well known that the energy dissipation within the electronic
system is so fast that on a time scale necessary for the lattice
to react (a few 10~ps) the local energy density is already too low
to create any permanent damage. As $d$ should not depend on the
different experimental conditions but only on the material we
plotted all the data (irradiations with Xe and Pb as well as
along the [100] and the [110] direction) in a single graph. As
one can see from fig.\ref{winkel} our data can be fitted quite
well with one curve, yielding $d=8$~nm.

\begin{figure}
\includegraphics[width=8cm]{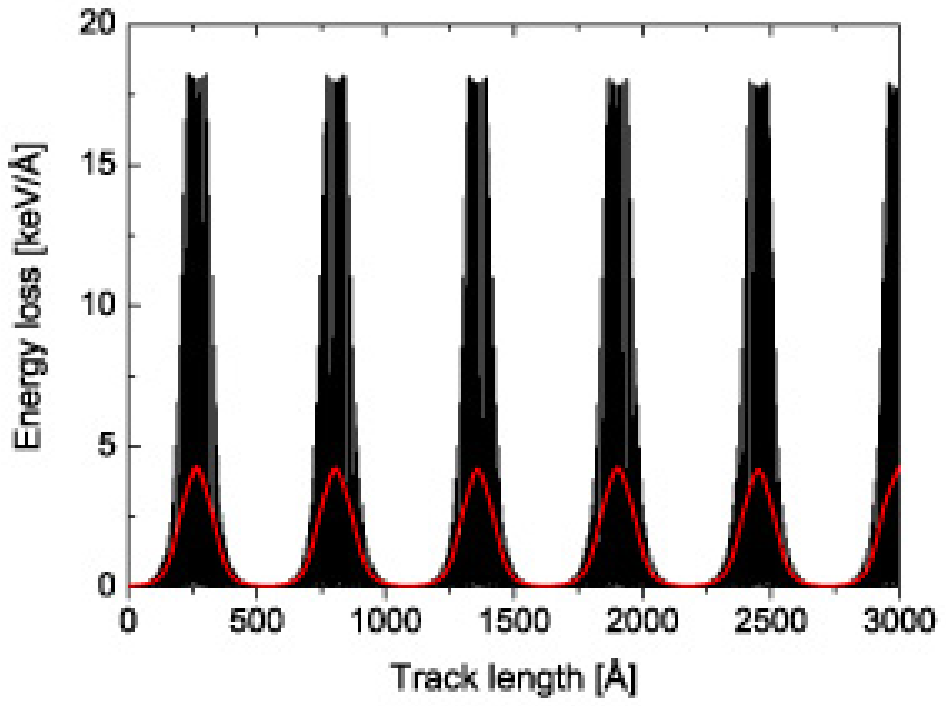}
\includegraphics[width=8cm]{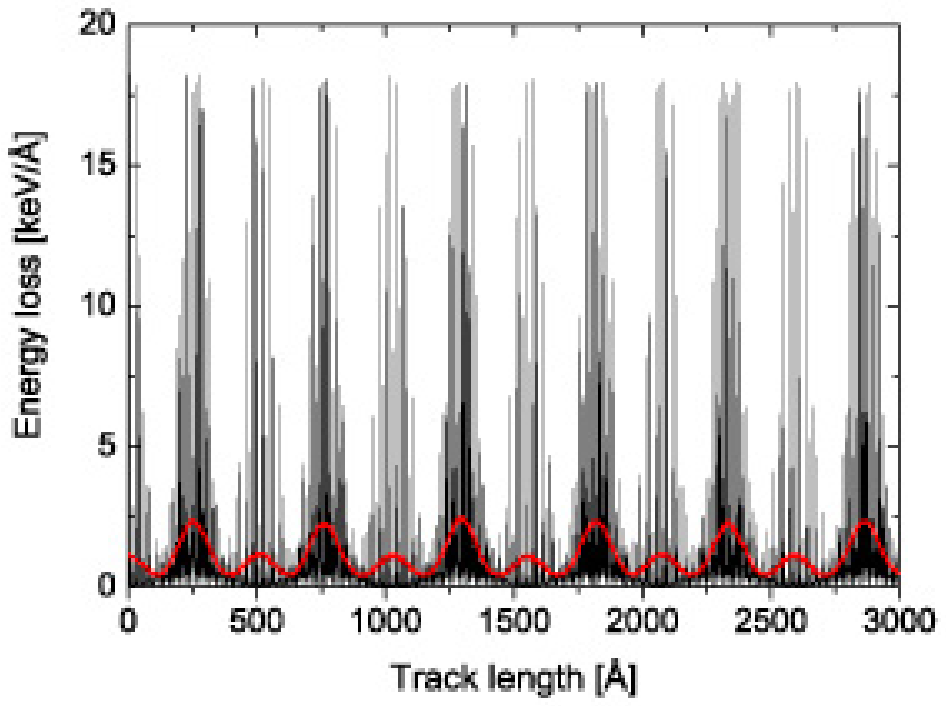}
\includegraphics[width=8cm]{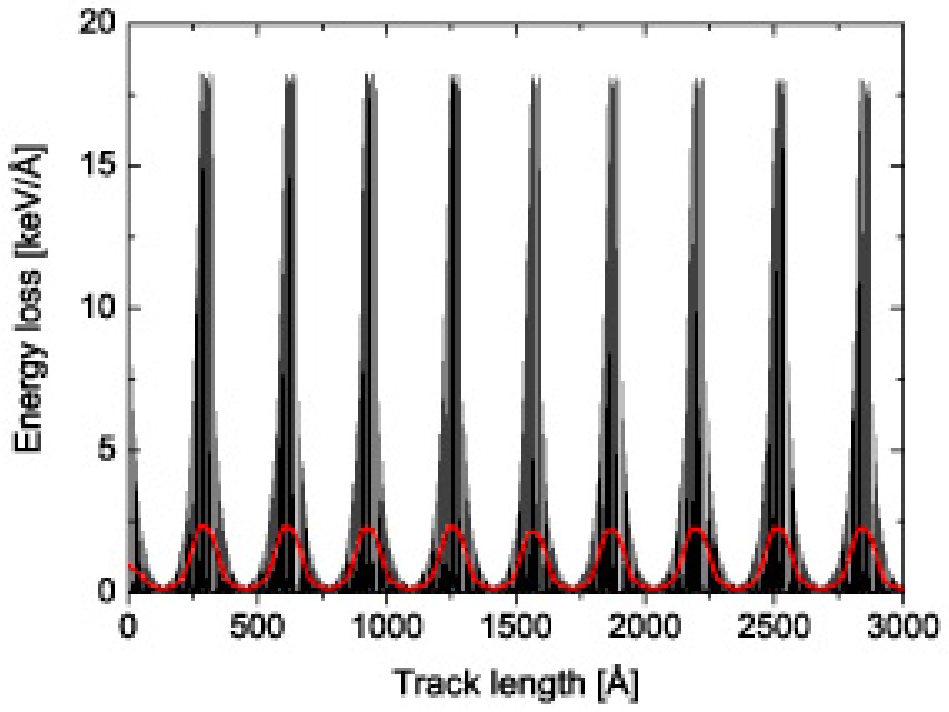}
\includegraphics[width=8cm]{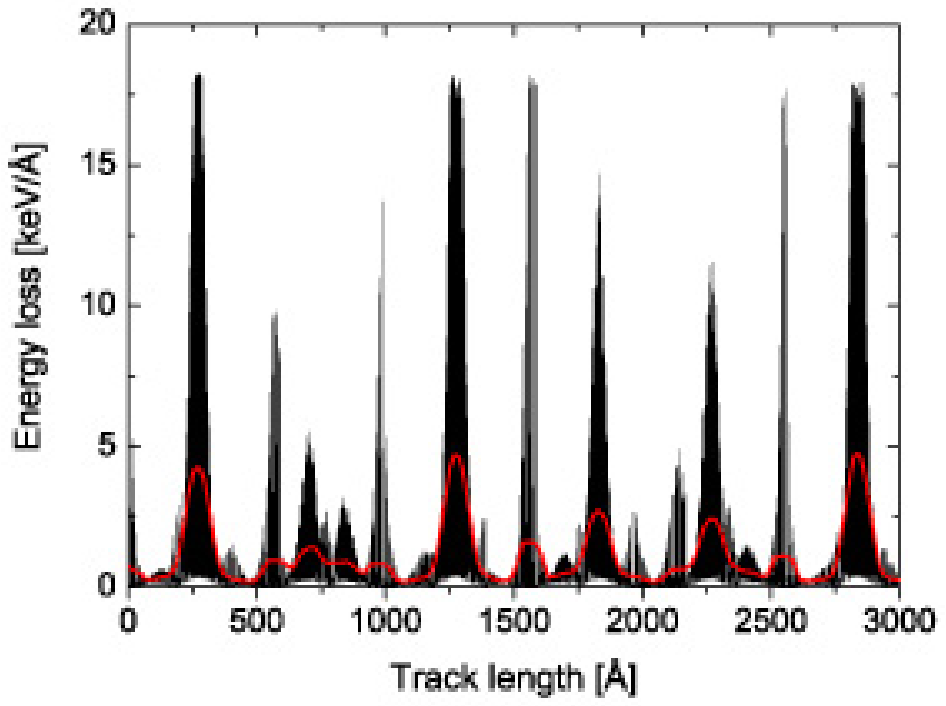}
\caption{Calculated electronic stopping along different example
trajectories with different $\theta$ and $\phi$. Top left panel:
$\Theta=0.4^\circ$ with respect to the [001] direction and $\phi$
along the [001] direction; top right panel: $\Theta=0.4^\circ$ and
$\phi=10^\circ$ with respect to the [001] direction. Lower left
panel: $\Theta=0.4^{\circ}$ with respect to the [011] direction
and $\phi$ along the [011] direction; lower right
panel:$\Theta=0.4^{\circ}$ with respect to the [011] direction
and $\phi=1.0^{\circ}$ with respect to the [011] direction. The
red line represents the moving average of the data, averaging
width 10~nm.} \label{stopping}
\end{figure}

Understanding the position of the individual hillocks within the
chain is more complicated and cannot be explained by this simple
relation. The reason is that the spatial details of the
electronic structure play an important role here. SrTiO$_3$ is an
insulator and most valence electrons are located close to the
oxygen atoms, a few around the titanium and almost none at the
strontium sites. The energy loss of the projectile is dominated
by electronic stopping which can only occur in regions where the
electron density is high enough. That is, the ion experiences a
strongly varying force corresponding to the local variations of
the electron density. We therefore follow the approach first
described in \cite{Akcoeltekin2007} and use the DFT electron
density \cite{DFT} to calculate the stopping power $dE/dx$ along
the ion trajectory \cite{Osmani2008-1}. The result for an ion
traveling almost (deviation $\approx 0,4^{\circ}$) along the
[001] and the [110] direction is shown in fig.\ref{stopping},
left part of top and lower panel, respectively.

We then calculate the moving average (averaging width 10~nm, red
line in fig.\ref{stopping}) of this data to take into account
that hillocks can only be created if the average energy density
in this region is high enough. The distance between the local
maxima in the averaged stopping power is 55~nm and correspond
well with the experimentally observed distances between hillocks.
However, we wish to point out that in certain geometries a slight
deviation from the azimuthal angle can have a rather large
influence on the exact positions of the maxima in the stopping
power along the trajectory. For example, a trajectory which
deviates only 1$^{\circ}$ from the low indexed [011] direction
towards the [112] direction (see fig.\ref{stopping}) appears
already very irregular. Therefore, the hillocks within the chains
are usually not as equally spaced as one would probably expect.
In addition, the electron excitation is a statistical process,
i.e. hillocks may appear differently even along identical
trajectories.

\begin{figure}
\includegraphics[width=8cm]{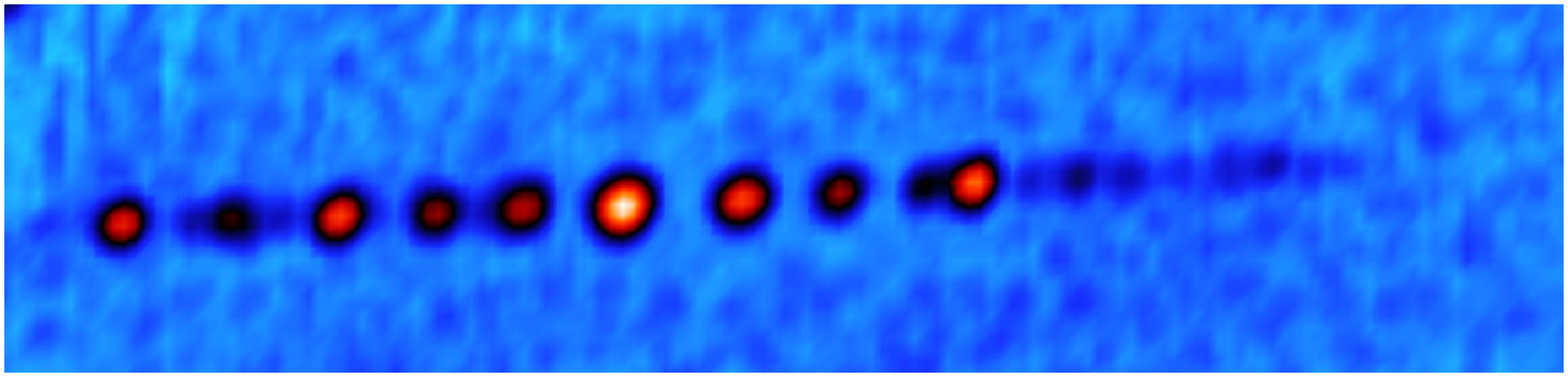}
\hspace{1cm}
\includegraphics[width=8cm]{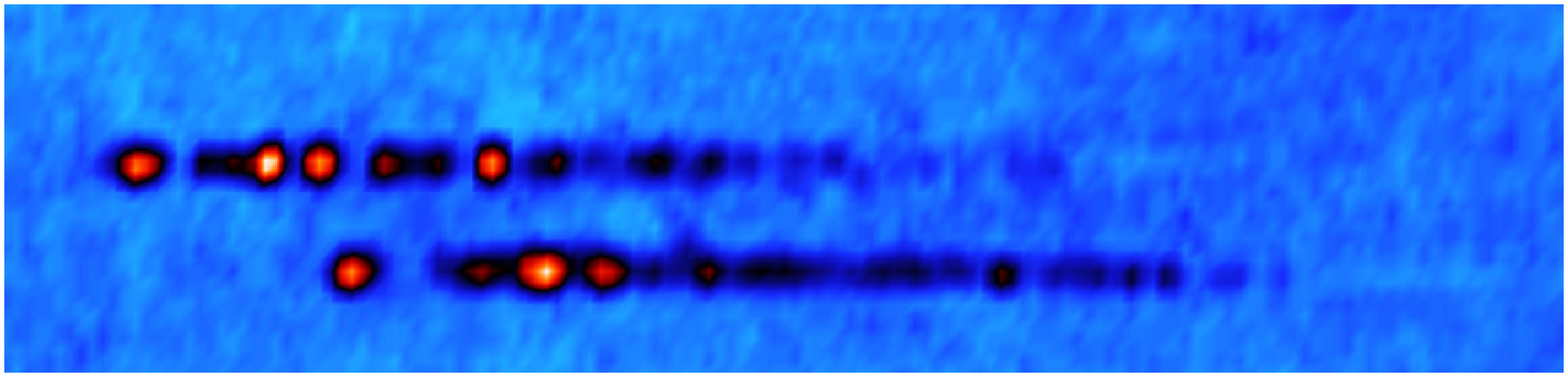}
\caption{Left: AFM image of a track created on a SrTiO$3$(100)
surface, $\theta=0.5^\circ, \phi\approx$ along the [001]
direction. Right: AFM image of a track created on a SrTiO$3$(111)
surface, $\theta=0.5^\circ, \phi\approx$ along the [110]
direction. Image size is $200 \times 850$~nm$^2$.}
\label{vgl100111}
\end{figure}

To demonstrate the influence of the azimuthal angle more clearly
we have irradiated a (100) oriented crystal surface along the
[001] direction and a (111) surface along the [110] direction
(exact within a few degrees). By using crystals of different
orientations it is possible to probe higher indexed directions
and still use glancing incidence angles. The result is shown in
fig.\ref{vgl100111}. The tracks appear very similar at first but
a detailed analysis reveals minor differences: On the (111)
surface the nanodots tend to appear in closely neighboring pairs
(right panel of fig.\ref{vgl100111}) whereas on the (100) surface
the dots appear to be more regularly spaced (left panel of
fig.\ref{vgl100111}, compare with top right panel of
fig.\ref{stopping}). In fact, the chain in the right part of
fig.\ref{vgl100111} exhibits a periodicity of $(58\pm8)$~nm,
which corresponds to a lattice constant of 3.9~\AA~ if we assume
that the ion hit the surface under $\Theta=0.4^\circ$. This
clearly indicates that the dots represent a direct projection of
the electron density onto the surface. Such a quantitative
comparison of any given trajectory with the calculations is
however not possible because the experimental parameters,
especially the angle of incidence, cannot be controlled with the
required precision.


From the analysis above it is evident that the electronic
excitation must be the source for the creation of nanodots.
However, the creation of dots requires the movement of atoms,
i.e.~the transfer of energy from the electronic into the phononic
system. Within the two-temperature model this is implemented by
introducing a electron-phonon-coupling constant $g$ which
controls the energy flux. As source terms the space and time
dependent energy loss along the trajectory enters the
calculation. In addition, the spatial and temporal evolution of
the energy density has to be calculated which is done by solving
the heat diffusion equation.

\begin{figure}
\includegraphics[width=10cm]{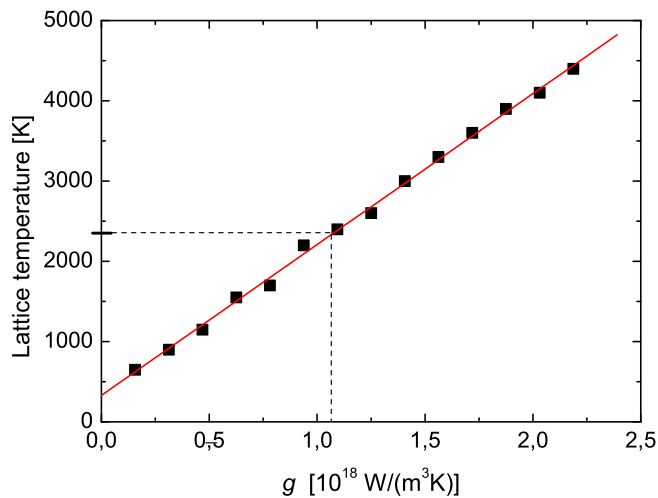}
\caption{Temperature in K on the surface as a function of the
electron-phonon coupling constant $g$. The bar on the temperature
axis marks the melting temperature of SrTiO$_3$. The red line
represents a linear fit to the data.} \label{kopplung}
\end{figure}

Since the parameter $d$ is material specific it can be used to
determine the electron-phonon-coupling constant as follows: from
fig.\ref{winkel} we can determine the track length that the ion
has to travel before it can no longer create modifications on the
surface. In the case of $\Theta=0.3^\circ$ and $d=8$~nm (see
\cite{Akcoeltekin2007} and fig.\ref{winkel}) this length would be
$\approx$~1530~nm. We calculate the energy loss at the end of that
trajectory and project it onto the surface. We subsequently
calculate the temperature at the surface for different values of
$g$. For the conversion of energy into temperature we use the
experimentally determined heat capacity of SrTiO$_3$ of
$C=100$~J/(K~mol)\cite{deLigny}. If we assume that in order to
modify the surface, at least the melting temperature of SrTiO$_3$
($T_{melt}=2353$~K \cite{handbook}) has to be achieved, we find
that $g\approx 1 \times 10^{18}$~W/(m$^3$~K) as can be seen from
fig.\ref{kopplung}. Experimentally determined values of $g$ for
other materials are in good agreement with our finding
\cite{Groenveld1995,Caron2006}.


After $g$ and the source terms $B(\vec{r},t)$ for the electronic
excitation have been determined the diffusion equations
describing the energy relaxation within the two temperature model
can be solved using finite differences and with von Neumann
boundary conditions. As initial temperature zero K was chosen.
\begin{equation}
\frac{\partial E_e}{\partial t}=D_e(\nabla E_e) - g \cdot (E_e -
E_l) + B(\vec{r},t)
\end{equation}
\begin{equation}
\frac{\partial E_l}{\partial t}=D_l(\nabla E_l) + g \cdot (E_e -
E_l)
\end{equation}

\begin{figure}
\includegraphics[width=12cm]{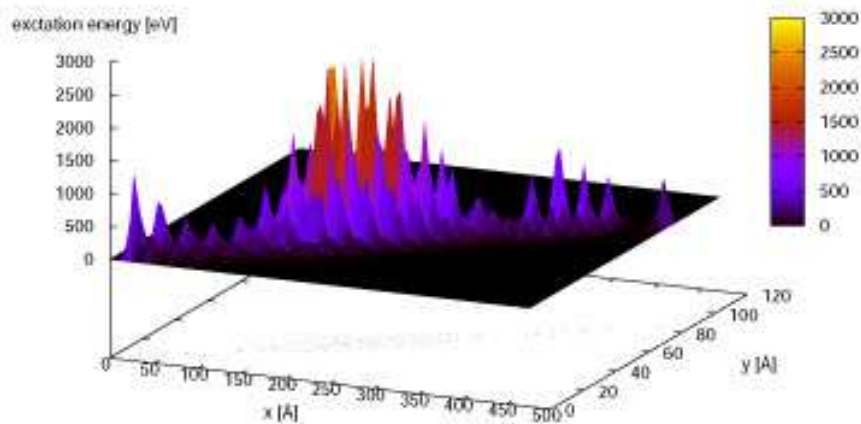}
\caption{Energy density on the surface for $g\approx 1 \times
10^{18}$~W/(m$^3$K) after $t=32$~fs. The diffusion of the energy
within the electronic system is treated with a constant diffusion
coefficient (see text).} \label{etempsim}
\end{figure}

Analyzing the data we find that with reasonable parameters
($D_l=0.0005$~cm$^2$/s, $D_e=0.05$~cm$^2$/s) the energy in the
electronic system dissipates on a time scale of a few ten
femtoseconds (see fig.\ref{etempsim}), whereas the energy
transfer into the phonon system requires picoseconds. That is,
the energy density that remains after 30~fs is much too low to
induce a temperature close to the melting temperature. To create
a high enough energy density in this scenario $g$ would have to
be three orders of magnitude larger which is non-physical.


If the energy transfer happens via electron-electron and
electron-phonon scattering the diffusion coefficient is directly
related to the scattering cross section of two particles. In the
case of highly excited electrons the scattering cross section
between a hot electron and a cold one is small because they are
energetically too different from each other. Only by transferring
energy to the phononic system the energy (and the temperature) of
the electrons decreases and thus the diffusivity increases. To
take this into account we use a temperature dependent diffusion
coefficient (see e.g.~\cite{Duvenbeck2005}) for the electronic
system instead of a constant diffusion. This approach treats the
excited electrons as an electron gas with a very low density
\cite{Baranov1988} (for a detailed discussion see
\cite{Osmani2008}):

\begin{equation}
D(T) = \frac{2k_B}{\pi m_e}\frac{T_e}{T_e^2+T_l}\label{eq:3}
\end{equation}
with $k_B$ the Boltzmann constant, $m_e$ the electron mass,
$T_e=\sqrt{2 E_e/C}$ and $T_l=2E_l/(3Nk_b)$.

\begin{figure}
\includegraphics[width=12cm]{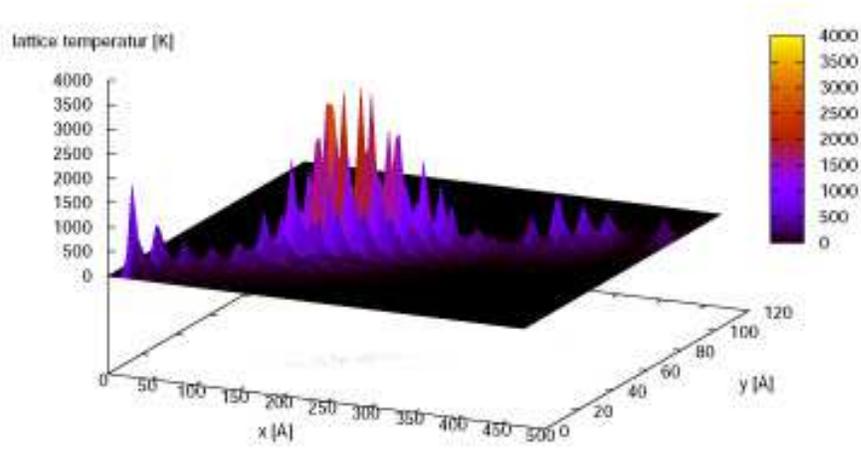}
\caption{Temperature on the surface in K for $g\approx 1 \times
10^{18}$W/(m$^3$K) after $t=34$~ps. The diffusion of the energy
within the electronic system is treated with a temperature
dependent diffusion coefficient (see text).} \label{tempsim}
\end{figure}

This approach ensures that the electronic dissipation is delayed.
As $D$ is proportional to $1/T$ for $T_l \ll T_e$ the diffusion is
very ineffective at high lattice temperatures. Thus, the
excitation is confined to a small region and lasts long enough to
feed sufficient energy into the much slower phonon system
characterized by a constant diffusion coefficient of
$D_l=0.0005$~cm$^2$/s. Solving the diffusion equations within this
approach, maxima of phonon energies are compatible with the
melting temperature and typically occur on time scales of
$\approx$ 50~ps (see fig.\ref{tempsim} and additional material),
which is quite reasonable.


That is, within the two-temperature model the time-scale problem
discussed above can be overcome by the introduction of a
temperature dependent diffusion coefficient. This does however
not solve the basic problem of the two-temperature model itself,
that the system is not in thermal equilibrium during the
excitation process and thus important physical quantities such as
the temperature are ill-defined. The conversion of energy into
electron temperatures e.g.~by using the heat capacity would only
be correct if we assume that the thermalization of the electronic
excitation happens either very fast or the distribution is
fermi-like from the beginning. For metals it has been shown that
the thermalization of a non-thermally excited electron gas
requires 10-100~fs \cite{Aeschlimann1994,Rethfeld2002}. In the
case of an insulator, this time should be even longer due to
longlived excitations such as excitons. In this sense, the
temperatures should be interpreted as a quantity to parameterize
the energy density but not as a real temperature.

Finally, we would like to discuss the possibility that the
creation of nanodots may not at all be the result of a thermal
process but is linked to a direct coherent excitation of atoms.
This so-called non-thermal melting has been observed for
semiconductors like InSb and is due to the efficient excitation
of optical phonons by intensive laser pulses \cite{Rousse2001}. A
prerequisite for this proccess is the excitation of a electron-
hole plasma by the intense laser pulse \cite{Stampfli1990}. With
ions, fields strength of similar intensity can be achieved and
thus a dense enough electron-hole plasma could be created. On the
basis of the current data it is not possible to check whether this
process exists. In any case it could not be treated within the
two-temperature model and is thus beyond the scope of this paper.

\section{Conclusions}
We have shown that with swift heavy ions chains of individual
hillocks can be created on SrTiO$_3$ surfaces if the irradiation
takes place under glancing angles. The length of the chains can be
easily controlled by varying the glancing angle. While the length
of the chains is independent of the azimuthal angle, the hillock
separation within the chains is not. This is a clear indication
that the exact spatial distribution of the electrons has to be
taken into account in order to understand the hillock formation.
We have demonstrated how within the two-temperature model the
electron phonon coupling constant can be obtained from the
dependence of the chain length on the angle of incidence.

Applying the two-temperature model to our data we find that the
electronic excitation dissipates too fast to deliver enough
energy to the phonon system. This problem can be overcome by using
a temperature-dependent diffusion coefficient, yielding
reasonable time scales for the phonon heating in the range of
several 10~ps. However, one has to keep in mind that only
relatively few electrons with rather high energies are created by
the moving ion. Thus, a model based on temperatures and
describing a diffusive transport might not be appropriate at all.
Instead, the transport would be better described applying a
Boltzman transport formalism which is part of the ongoing
research in our group.

\section*{Acknowledgement}
We thank R. Meyer for calculating the electron density and for
stimulating discussions. We thank A. Duvenbeck for many helpful
discussions and his continuous support. Financial support by the
DFG - SFB616: {\it Energy dissipation at surfaces} is gratefully
acknowledged. The experiments were performed at the IRRSUD
beamline of the Grand Accelerateur National d'Ions Lourds (GANIL),
Caen, France.


\section*{References}
\bibliography{STO}
\bibliographystyle{unsrt}

\clearpage

\clearpage

\end{document}